\documentclass[nofootinbib]{revtex4}
\usepackage{amsfonts,amssymb,amsmath,amsthm,amscd}
\usepackage{bm}
\usepackage{graphicx}
\usepackage{dcolumn}
\usepackage{hyperref}
\usepackage{enumerate}
\def\demi{\frac{1}{2} } 
 \newcommand{\dr}{\rightarrow}

    \def\Ee{{\cal E}}  %\def\gg{{\cal G}}
     \def\mm{{\cal M}} \def\nn{{\nonumber}}
      
  \def\xx{{\cal X}} \def\yy{{\cal Y}} \def\zz{{\cal Z}}

\newcommand{\be}{\begin{equation}}
\newcommand{\bee}{\begin{equation}}
\newcommand{\eeq}{\end{equation}}

\def\mmm{\mathfrak{m}}
\def\nnn{\mathfrak{n}}

\newcommand{\ket}[1]{|#1\rangle} \newcommand{\bra}[1]{\langle #1|}

 \newcommand{\bem}{\begin{bmatrix}} \newcommand{\eem}{\end{bmatrix}}
\newcommand{\beq}{\begin{equation}} \newcommand{\ee}{\end{equation}} \newcommand{\eq}{\end{equation}}
\newcommand{\beqa}{\begin{eqnarray}} \newcommand{\eeqa}{\end{eqnarray}}

\def\ie{{i.e. \hspace{-.05cm}}}

\def\cf{{c.f. \hspace{-.05cm}}}

\def\mmm{{\mathfrak{m}}}

\def\xx{{\mathrm{x}}}
\def\yy{{\mathrm{y}}}
\def\zz{{\mathrm{z}}}

\newcommand{\hpsi}{\hat{\Psi}}
\newcommand{\hchi}{\hat{\chi}}
\newcommand{\ha}{\hat{a}}
\newcommand{\hab}{\hat{b}}
\newcommand{\hnew}{\lambda}
\newcommand{\habd}{\hat{b}^{\dagger}}
\newcommand{\had}{\hat{a}^{\dagger}}
\newcommand{\phigrav}{\Phi_{\mathrm{grav}}}
\newcommand{\csou}{c_{s}}
\newcommand{\gloc}{G_{N}^{loc}}
\newcommand{\hphi}{\hat{\phi}}
\newcommand{\kkk}{{\mathrm{k}}}
\newcommand{\hhh}{{\mathrm{h}}}
\newcommand{\pp}{{\mathrm{p}}}
\newcommand{\dv}{\mathrm{d}^{3}}

\begin{document}

\title{Gravitational dynamics in Bose Einstein condensates }
\author{F.~Girelli\footnote{girelli@sissa.it}, S.~Liberati\footnote{liberati@sissa.it}, L.~Sindoni\footnote{sindoni@sissa.it},  } \affiliation{SISSA and INFN, Sezione di Trieste \\
Via Beirut 2-4, 34014 Trieste, Italy}
\date{\today}
\begin{abstract}
Analogue models for gravity intend to provide a framework where  matter and gravity, as well as their intertwined dynamics, emerge from degrees of freedom that have a priori nothing to do with what we call gravity or matter. Bose Einstein condensates (BEC) are a natural example of analogue model since one can  identify matter  propagating on a (pseudo-Riemannian) metric with collective excitations above the condensate of atoms. However, until now, a description of the ``analogue gravitational dynamics" for such model was missing. We show here that in a BEC system {with massive quasi-particles}, the gravitational dynamics {can be} encoded in a modified (semi-classical) Poisson equation. In particular, gravity is of extreme short range (characterized by the healing length) and the cosmological constant appears from the non-condensed fraction of atoms in the quasi-{particle} vacuum. {While some of these features make the analogue gravitational dynamics of our BEC system quite different from standard Newtonian gravity, we nonetheless show that it can be used to draw some interesting lessons about ``emergent gravity" scenarios.}
\end{abstract}

\maketitle

\section*{Introduction}

Analogue models for gravity have provided a powerful tool for testing (at least in principle) kinematical features of classical and quantum field theories in curved spacetimes \cite{Barcelo:2005fc}. The typical setting is the one of sound waves propagating in a perfect fluid \cite{unruh,visseracou}. Under certain conditions, their equation can be put in the form of a Klein-Gordon equation for a massless particle in curved spacetime, whose geometry is specified by the acoustic metric. {Among the various condensed matter systems so far considered, Bose-Einstein condensate (BEC) \cite{Garay:1999sk,Barcelo:2000tg} had in recent years a prominent role for their simplicity as well as for the high degree of sophistication achieved by current experiments. In a BEC system one can consider explicitly the quantum field theory of the quasi-particles (or phonons), the massless excitations over the condensate state, propagating over the condensate as the analogue of a quantum field theory of a scalar field propagating over a curved effective spacetime described by the acoustic metric. It provides therefore a natural framework to explore different aspects of quantum field theory in various interesting curved backgrounds (for example quantum aspects of black hole physics \cite{Balbinot:2004da, BLVsemicls collapse} or the analogue of the creation of cosmological perturbations \cite{BLVFRW, Silke-Infla}).
Unfortunately, up to now, the analogy with gravity is only partial: there is no analogy with some sort of (semiclassical) Einstein equations, since it has not been possible to put the fluid equations, which are those describing the dynamics of the acoustic metric, in a geometrical form which could eventually lead to a complete dynamical analogy with general relativity \cite{Barcelo:2001tb}.  Our aim is to fill this gap {and to gather from the description of the gravitational dynamics general lessons about possible features of ``emergent gravity" scenarios.}

In BEC, the effective emerging metric depends on the properties of the condensate wave-function. One can expect therefore the gravitational degrees of freedom to be encoded in the variables describing the condensate wave-function \cite{Barcelo:2000tg}, which is solution of the well known Bogoliubov--de Gennes (BdG) equation. The dynamics of gravitational degrees of freedom should then be  inferred from this  equation, which is essentially non-relativistic.  The gravitodynamics of the BEC should therefore be the analogue of some sort of {Newtonian gravity}, and we shall reinterpret the BdG equation as a modified Poisson equation.

The ``emerging matter", the quasi-particles, in the standard BEC, are phonons, \ie~massless excitations. Since we expect the quasi-particles to be the matter source in the Poisson equation, we run a priori into a problem:  massless particles are not treatable in the framework of Newtonian mechanics. To avoid this issue, we shall then introduce a new term in the BEC Hamiltonian  which will softly break the usual $U(1)$ symmetry and therefore will allow the quasi-particles to acquire mass.
}

The plan of the paper is as follows. In the first section, we discuss the dynamics of the condensate and of the quasi-particles in the particular case of a BEC with a soft $U(1)$ symmetry breaking term. We show how this modification affects the dynamics of the condensate, leading to a generalized BdG equation, and how it provides a mass to the quasi-particles.
In the second section, we focus on identifying the analogue of the gravitational potential and we show that it has to be identified with inhomogeneities in the condensate density. We then show that the BdG equation can be cast in a form that closely mimic a Poisson equation. By looking at different sources we shall then identify the analogue of the Newton constant (and its possible dependence on the momentum of the quasi-particles) as well as the cosmological constant. The final outcome of our analysis is that the  gravitodynamics of the BEC  with massive quasi-particles is encoded in a modified Poisson equation characterized by two main features. First gravity is of very short range, \ie a localized particle generates a  Yukawa type potential.  Secondly the cosmological constant naturally appears from the non-condensed fraction of atoms in the quasi-particle vacuum.

\section{A modified BEC}\label{sec:becsetup}

Let us start recalling the standard setup for a BEC of many atoms in a box of volume $V$, in the dilute gas approximation.
In this limit it is possible to describe the  atoms via a second-quantized field operator
\begin{equation}
\hpsi = \frac{1}{\sqrt{V}} \sum_{k} \ha_{k} e^{ik\cdot \xx},
\end{equation}
whose evolution is encoded in the  Hamiltonian $\hat H_0$  \cite{blackbook}
 \begin{equation}\label{BEC Hamiltonian}
\hat{H}_{0} = \int \hpsi^{\dagger}(\xx)\left( -\frac{\hbar}{2m} \nabla^{2}- \mu + \frac{\kappa}{2} |\hpsi|^{2} \right) \hpsi(\xx) \dv\xx,
\end{equation}
which generates the standard non-linear  equation\footnote{The field operator $\hpsi$ is obeying the commutation relations:
$$
[\hpsi(\xx),\hpsi^{\dagger}(\yy)] = \delta^{3}(\xx-\yy),\qquad [\hpsi(\xx),\hpsi(\yy)] =[\hpsi^{\dagger}(\xx),\hpsi^{\dagger}(\yy)] =0.
$$}
\begin{equation}\label{eq:fieldeq0}
i\hbar \frac{\partial }{\partial t}\hpsi = [\hat H_0, \hpsi] =  -\frac{\hbar^2}{2m} \nabla^{2}\hpsi - \mu \hpsi + \kappa |\hpsi|^{2}\hpsi.
\end{equation}
The operator $\hpsi$ has  dimension  $L^{-3/2}$: the quantum average of its modulus square on a given state represents the number density of atoms. The mass $m$ is the mass of the atoms. The energy $\mu$ is the chemical potential and has dimension $ML^{2}T^{-2}$. The constant $\kappa$, of dimension $ML^{5}T^{-2}$, represents the strength of the two-bodies interaction between atoms. We neglect higher order contributions to the second quantized Hamiltonian.

{While the condensation process can be easily understood as a macroscopic occupation number of an energy level, there are several approaches to describe it mathematically. The mean field approach is particularly convenient: we say that the system of $N$ bosons has condensed whenever the field $\hpsi$ develops a non-zero vacuum expectation value (vev)
\begin{equation}
\bra{\Omega} \hpsi \ket{\Omega} = \psi,
\end{equation}
where $\psi$ is  a classical complex field describing the condensate and it is sometimes called the condensate wave-function. If this mean field is non-vanishing, we have that the two point correlation function
\begin{equation}
 G(\xx,\yy) = \langle \hpsi^{\dagger}(\xx) \hpsi(\yy) \rangle,
\end{equation}
tends to a non-zero constant when $\xx,\yy$ are infinitely separated, \ie~the system develops long range correlations \cite{blackbook}.

However, the mean field approximation is not the most rigorous method to treat a BEC system. In particular, the particle-number-conserving
approach %\cite{castindum} 
has been proven to give more accurate predictions for the physical properties of the condensate in settings in which the number of atoms, $N$, is fixed. This dimensionless quantity, then, is used to expand systematically the equation for the evolution of the operator $\hpsi$, in powers of $N^{1/2}$. While this method was shown to provide a more accurate description of BEC systems, it has also shown that the mean field approximation gives already very good predictions for quasi-static configurations~\cite{castindum}.

In this paper, we consider an idealized case where the confining potential is almost constant, both in space and in time. Moreover, we neglect boundary effects due to the finite size of the trap, and we assume that all the other experimental parameters are weakly time-dependent. In this case, then, the mean field approximation is well motivated.

In order to derive the properties of the BEC, we split the field operator in the condensate part $\psi$ (the mean field) and an operator $\hchi$ representing the atoms which are not in the condensate:
\begin{equation} \label{condensation}
\hpsi = \psi \mathbb{I} +\hchi,
\end{equation}
and use this form of the operator in the evolution equation \eqref{eq:fieldeq0}.}

{For the reasons discussed in the Introduction and further analyzed in section} \ref{sec:grav}, we need to slightly modify the standard Hamiltonian $H_0$
by introducing a term which is (softly) breaking the $U(1)$ symmetry in
\eqref{eq:fieldeq0}.
\beq \hat H_0 \dr  \hat H = \hat H_0+ \hat H_\lambda, \qquad \hat H_\lambda = - \frac{\lambda}{2} \int \dv\xx \left( \hpsi(\xx) \hpsi(\xx) + \hpsi^{\dagger}(\xx) \hpsi^{\dagger}(\xx) \right).\label{sy-br} \ee
The parameter $\lambda$  has the same dimension as $\mu$. With this new Hamiltonian, the non-linear equation \eqref{eq:fieldeq0} becomes
\begin{equation}\label{eq:fieldeq}
i\hbar \frac{\partial }{\partial t}\hpsi = [\hat H , \hpsi] =  -\frac{\hbar^2}{2m} \nabla^{2}\hpsi - \mu \hpsi  + \kappa |\hpsi|^{2}\hpsi- \lambda \hpsi^{\dagger}.
\end{equation}
We shall show in section \eqref{sec:quasipart} that $\hat H_\lambda$ generates  a mass for the quasi-particle\footnote{If we were discussing a relativistic field theory this massive particle would be nothing else than a pseudo-Goldstone boson.}.  Even though $\hat H_\lambda$  both creates and destroys pairs of atoms, it is not difficult to check that $\hat H_\lambda$ is not commuting with the number operator $\hat N$,
\begin{equation}\label{eq:commutatorHN}
[\hat{H}_{\lambda},\hat{N}] =
-\lambda \int \dv \xx \left( \hpsi(\xx)\hpsi({\xx}) -\hpsi^{\dagger}(\xx)\hpsi^{\dagger}(\xx)\right)
\end{equation} while unitarity is preserved. In fact, when  applied on a state with a definite number of atoms $n$ we have:
\begin{equation}\label{eq:number}
\ket{n} \rightarrow \ket{n-2} + \ket{n+2},
\end{equation}
which means that an eigenstate of the number operator is taken into \emph{a superposition} of states with different occupation numbers\footnote{As a consequence, the particle-number-approach cannot be used as it is defined in \cite{castindum}. A more general treatment should be developed in order to include these interesting situations where the number of atoms is not a conserved operator. This extension could also provide a test for the accuracy of the mean field approximation, and hence on the stability of the condensation, in the U(1) breaking case.}.
However, the expectation value of the number of operator on its eigenstates is still constant
\begin{equation}
i \hbar \frac{\partial}{\partial t} \bra{n}\hat{N} \ket{n} =  \bra{n}[\hat{N},\hat{H}]\ket{n} = \bra{n}[\hat{N},\hat{H}_{\hnew}]\ket{n}  \propto \langle n | n-2 \rangle -\langle n | n+2 \rangle =0.
\label{eq:Nav}
\end{equation}

Finally, we want to discuss some plausible physical setup corresponding to the $U(1)$ symmetry breaking in BECs. Given the relationship existing between the Hamiltonian \eqref{sy-br} and the number operator, a natural situation would be represented by a condensate which is able to exchange particles with some sort of reservoir, in such a way to preserve, on average, their number. Several settings in this sense could be conceived, {\em e.g.} with coupling with suitably tuned lasers. Another interesting concrete example is represented by the case of magnon condensates, see \cite{magnon} and references therein.

\subsection{Condensate}\label{sec:cond}
We consider the dynamics generated by \eqref{eq:fieldeq}, from which we want {to} extract the equation of motion  for the condensate $\psi$.  The evolution of the mean field $\psi$ is easily determined in terms of the eigenstates $\ket{E}$ of the Hamiltonian $\hat H$:
\begin{equation}\label{BdG}
i\hbar \frac{\partial}{\partial t} \psi =i\hbar \frac{\partial}{\partial t } (\bra{E} \hpsi \ket{E})
=\bra{E} i\hbar \frac{\partial}{\partial t} \hpsi \ket{E} =
-\frac{\hbar^{2}}{2m } \nabla^{2} \psi - \mu \psi -\lambda \psi^{*} + \kappa |\psi|^{2} \psi + 2 \kappa \nnn_{E} \psi + \kappa \mmm_{E} \psi^{*},
\end{equation}
where $\mmm_{E}=\bra{E}\hchi^{2}\ket{E}, \nnn_{E}=\bra{E}\hchi^{\dagger}\hchi\ket{E}$ encode the effect of the non-condensate atoms.
This is the generalization of the Bogoliubov-de Gennes (BdG) equation for the condensate wave-function to the case $\lambda\neq 0$.

If we have $N$ particles~\footnote{Throughout the paper we shall intend by $N$ the average number of particles which is conserved (cf.~equation~\eqref{eq:Nav}).} in the condensate, the number density of the non-condensate fraction is of order $1/N$ with respect to the number density of the condensate. In particular, the terms $\mmm,\nnn$ are of order $1/N$.
At zeroth order, we have   the generalization of the Gross-Pitaevskii (GP) equation:
\begin{equation}\label{eq:GP}
i\hbar \frac{\partial}{\partial t} \psi = -\frac{\hbar^{2}}{2m } \nabla^{2} \psi - \mu \psi -\lambda \psi^{*} + \kappa |\psi|^{2} \psi.
\end{equation}
The time independent homogeneous solution to the GP equation is
\begin{equation}\label{eq:condensatedens}
n_{c} = |\psi|^{2} = \frac{\mu+\lambda}{\kappa},
\end{equation}
where we have fixed the phase of the condensate to be zero\footnote{In appendix \ref{app:fluid} we show that this is not an arbitrary choice, but rather a consequence of the situation we want to describe.}.

We define the \emph{healing length} $\xi$ as the length scale at which the kinetic term is of the order of the self-interaction term:
\begin{equation}\label{healing}
\frac{\hbar^{2}}{2m \xi^{2}} = \kappa n_{c} \Leftrightarrow \xi^{2} = \frac{\hbar^{2}}{2m \kappa n_{c}}.
\end{equation}

This length represents the spatial scale needed for the condensate to pass from the value $n_c=0$ at the boundary of the region where it is confined to the bulk value $n_c$. In other words it represents the scale of the dynamical processes involving the deformation of the condensate wavefunction.

\subsection{Quasi-particles}\label{sec:quasipart}
The equation of motion for the particles out of the condensate is obtained by subtracting the equation for the condensate \eqref{eq:GP} from the equation for $\hpsi$  given in \eqref{eq:fieldeq}.
We are interested in the propagating modes, so we neglect the self-interactions. We obtain:
\begin{equation}\label{dynamics for quasiparticles}
i\hbar\frac{\partial}{\partial t} \hchi = -\frac{\hbar^{2}}{2m }\nabla^{2} \hchi + (2 \kappa |\psi|^{2}-\mu) \hchi + (\kappa \psi^{2}-\lambda) \hchi^{\dagger}.
\end{equation}
Let us consider the case of homogeneous condensate with density $n_{c}$ given above.
In this situation we have:
\begin{equation}
i\hbar\frac{\partial}{\partial t} \hchi = -\frac{\hbar^{2}}{2m }\nabla^{2} \hchi + (\mu+2\lambda) \hchi + \mu \hchi^{\dagger}.
\end{equation}
If we decompose the field $\hchi$ in its plane wave components, we can rewrite this equation as
\begin{equation}
i\hbar\frac{\partial}{\partial t} \ha_{\kkk} = \frac{\hbar^{2} k^{2}}{2m} \ha_{\kkk} + (\mu+2\lambda)\ha_{\kkk} + \mu \had_{-\kkk}.
\end{equation}
The mixing between $\ha$ and $\had$ due to the evolution in time becomes then apparent. We therefore pass to the quasi-particle operators $\hphi(\xx)$
\begin{equation}
\hphi(\xx) = \frac{1}{\sqrt{V}} \sum_{\kkk} \hat{b}_{\kkk} e^{i\kkk\cdot \xx},
\end{equation}
which are related to the particle operators through the Bogoliubov transformation
\begin{equation}\label{bogolobov transf}
 \ha_{\kkk} = \alpha(\kkk) \hat{b}_{\kkk} + \beta(\kkk)\hat{b}_{-\kkk}, \quad \textrm{with } \alpha^2(k)-\beta^2(k)=1.
\end{equation}
 $\alpha,\beta$ are only functions of $k=|\vec\kkk|$, since the condensate is homogeneous and isotropic. The equation of evolution for the quasi-particles is then given by
\begin{equation}
 i\hbar \frac{\partial }{\partial t} \hat{b}_{\kkk} = \mathcal{E}(k)\hat{b}_{\kkk},
\end{equation}
with the energy
\begin{equation}\label{eq:qparticledr}
\mathcal{E}(k) = \left( \frac{\hbar^{4} k^{4}}{4m^{2}} + 4\lambda (\mu+\lambda) + \frac{\mu+2\lambda}{m} \hbar^{2}k^{2}  \right)^{1/2}.
\end{equation}
The Bogoliubov coefficients are given by:
\begin{equation}\label{Bogoliubovcoeff}
\alpha^{2}(k) = \frac{A(k)+\mathcal{E}(k)}{2 \mathcal{E}(k)},
 \qquad \beta^{2}(k) = \frac{1}{2\mathcal{E}(k)}\frac{\mu^2}{A(k)+\mathcal{E}(k)},
\end{equation}
where we have introduced the quantity
\begin{equation}
 A(k) = \frac{\hbar^{2}k^2}{2m} + \mu + 2\lambda.
\end{equation}
The high energy limit of these coefficients is:
\begin{equation}
\lim_{k \rightarrow \infty} \alpha^{2}(k) = 1, \qquad \lim_{k\rightarrow \infty} \beta^{2}(k)= 0,
\end{equation}
which means that  at large wave-number (and hence large momentum), the quasi-particle operators coincide with the particle operators. This matches the behavior of the energy, which becomes just the energy of a non-relativistic particle of mass $m$, just like a free atom.
The dispersion relation \eqref{eq:qparticledr} suggests the introduction of the following quantities:
\begin{equation} \label{eq:soundspeedmass}
\csou^{2} =\frac{\mu + 2 \lambda}{m},  \qquad \mathcal{M}^{2} = 4\frac{\lambda(\mu+\lambda)}{(\mu+2\lambda)^{2}}m^{2}.
\end{equation}
$\csou$ plays the role of the speed of sound, while $\mathcal{M}$ plays the role of a rest mass for the quasi-particle. Since $\mm$ is proportional to $\lambda$, we clearly see that it is the  term $\hat H_\lambda$ that generates the mass of the quasi-particle. When  $\lambda\dr0$, that is when $\hat H\dr \hat H_0$, the quasi-particle becomes massless, \ie a phonon, and the speed of sound reduces to the usual one in BEC.

Notice that, in order to have a non-negative mass square, we have to require $\lambda\geq0$. In standard BEC, we usually assume that the chemical potential $\mu$ is positive: indeed if it was not positive, there couldn't be any condensation. In our case, we can relax this requirement and obtain that $\mu>-\lambda$ as a condition. In the following we consider $\mu>0$, in order to be able to consider the case in which the correction we are inserting is very small, without affecting dramatically the condensation. {Indeed, it is easy to see that a condensation can take place in the system with a soft $U(1)$ breaking by checking the behavior of the two points correlation function $G(\xx,\yy)=\langle \hpsi^{\dagger}(\xx) \hpsi(\yy)\rangle$, which is still showing the presence of long range correlations, since the mean field $\psi$ is non-vanishing (cf. equation \eqref{eq:condensatedens}).}

$\mm$ is proportional to $m$, the mass of the atoms. By introducing the ratio $\zeta= \lambda/\mu$, we  introduce the function $F(\zeta)$
\begin{equation}
\mm^2= F(\zeta)m^2=4 \frac{\zeta(1+\zeta)}{(1+2\zeta)^{2}}m^2.
\end{equation}
Under our assumptions, we have that $\zeta \geq 0$. It is then straightforward to check that on this domain $F(\zeta)$ is a monotonic (increasing) function and that
\begin{equation}
F(0)=0, \qquad \lim_{\zeta\rightarrow + \infty}F(\zeta) = 1.\nn
\end{equation}
We conclude therefore  that the mass of the quasi-particles $\mm$ is always bounded by the mass of the atoms, $\mm\in [0,m)$.

It is also interesting to notice that using the variable $\zeta$, the speed of sound is:
\begin{equation}
\csou^{2} = \frac{1 + 2\zeta}{1+\zeta} \frac{\kappa n_{c}}{m}.
\end{equation}
For $\zeta$ small, we then have $\csou^{2} \approx \kappa n_{c}/m$, which is the standard result, while, for $\zeta\rightarrow \infty$, $\csou^{2}\rightarrow 2 \kappa n_{c}/m$.
\subsection{The various regimes for the MDR} \label{sec:regimes}
Before moving on to the gravitational dynamics, let us  discuss briefly the content of the dispersion relation  \eqref{eq:qparticledr} for the quasi-particles, rewritten using $\csou$ and $\mm$.
\begin{equation}\label{eq:qparticledr2}
\mathcal{E}(p) = \left(\frac{p^{4}}{4m^{2}} + \csou^{2} p^{2} + \mm^{2}\csou^{4}\right)^{1/2},
\end{equation}
where we are using the obvious notation $p=\hbar k$ to simplify the shape of the expressions.
Let us define the  characteristic momenta $p_A$, $p_B$ and $p_C$ such that
\begin{equation}
\frac{p_{A}^{4}}{4m^{2}} = \csou^{2} p_{A}^{2}, \qquad \frac{p_{B}^{4}}{4m^{2}} = \mm^{2} \csou^{4},
\qquad \csou^{2} p_{C}^{2} = \mm^{2} \csou^{4}, \nn
\end{equation}
so that they are explicitly
\begin{equation}
p_{A}^{2} = 4 m^{2} \csou^{2}, \qquad p_{B}^{2} = 2 m \mm \csou^{2},
\qquad p_{C}^2=\mm^{2} \csou^{2}.\nn
\end{equation}
They are related through the relations
\begin{equation}
p_{C}^{2} = 2 F(\zeta) p_{B} ^{2}=
4 F^{2}(\zeta) p_{A}^{2}.\nn
\end{equation}
If $\zeta \ll 1$, which will be the regime we shall consider,   we have also that
\begin{equation}\label{regimes}
p_{C}\ll p_B  \ll p_{A}.
\end{equation}
Taking into account \eqref{regimes}, the  characteristic momenta define different regimes:
\begin{itemize}
\item If $p \gg p_A$, the  term $p^4$ dominates, the dispersion relation \eqref{eq:qparticledr2} is well approximated by $\mathcal{E} \sim p^2/2m$, we are in the transphononic regime.

\item If on the contrary we have $p_C \ll p\ll p_A$, we can safely neglect the term of order $p^4$, we are then in the relativistic regime since the dispersion relation \eqref{eq:qparticledr2} is well approximated by $\mathcal{E} \sim (p^2\csou^2 + \mm^2\csou^4)^\demi$. The quasi-particle is then relativistic, when the speed of sound $c_{s}$ is playing the role of the speed of light.
\item If we are in the regime where $p\ll p_{C}$, this means that the quasiparticle has a speed much smaller than $\csou$, so that this is the Galilean limit of the relativistic regime. We are then dealing with a Galilean quasi-particle. The rest mass $\mm \csou^2$ provides the usual constant shift of the Galilean energy $\Ee\sim \mm \csou^2 + \demi p^2/\mm  $.
\end{itemize}

\section{Gravitational dynamics}\label{sec:grav}
Since we have described the physics of the system in the case of homogeneous condensate $\psi$, we can now pass to study the inhomogeneous one, and hence the emergence of a gravitational dynamics. We are going to consider  condensates which are nearly, but not exactly, homogeneous:  this will correspond to the case of weak gravitational field.

In a curved spacetime, the identification of the Newtonian gravitational potential requires a non-relativistic limit of the geodesic equation in a weak gravitational field  \cite{Misner:1974qy}. For instance, in asymptotically flat spacetimes, there is a coordinate system such that the metric, in the asymptotic region, can be written as $g_{\mu\nu}=\eta_{\mu\nu}+h_{\mu\nu}$, where $h_{\mu\nu}$ encodes the deviation from flatness, \ie the gravitational field. In this regime, it is possible to show that the Newtonian gravitational field is identified with the component $h_{00}$:
\begin{equation}
\Phi_{N}(\xx) = -\frac{1}{2} h_{00}(\xx)
\end{equation}
It is well known that in the context of standard BEC (\ie dealing with the non-linear equation \eqref{eq:fieldeq0}), the quasi-particles travel in an emergent metric $ds^2$ determined in terms of the homogenous condensate $\psi$ \cite{Barcelo:2000tg}.
\begin{equation}\label{eq:acousticlineelementasym}
ds^{2} =\frac{n_{c}}{m c_{s}} \left[-\left(c^{2}_{s}- v^{2} \right) dt^{2} -2 v_{i} dt d\xx^{i} +\delta_{ij} d\xx^{i}d\xx^{j} \right],
\end{equation}
where $m$ is the mass of the atoms and $c_s$ and $\vec v$  depend on the properties of the condensate $\psi=\sqrt{n_c}e^{i\theta}$, through $$c_c=\frac{\kappa n_c}{m}, \qquad \vec v=\frac{1}{m}\vec \nabla \theta.$$
Considering that the condensate is homogenous, the density and velocity   profiles become constant, \ie respectively   $n_c=n_\infty$, $\vec v=\vec v_{\infty}$.
With the coordinate transformation,
\begin{equation} \label{eq:pgtominkowski}
dT=dt, \qquad dX^{i} = d\xx^{i} - v^{i}_{\infty}dt,
\end{equation}
the line element \eqref{eq:acousticlineelementasym} is rewritten as:
\begin{equation}
ds^{2}_{\infty} = -c_{\infty}^{2} dT^{2} + d\bf{X}^{2}.
\end{equation}
The condition of asymptotic flatness for spacetimes can be translated with the condition of asymptotic homogeneity for the condensate.  We require then that only in a small region of space, in the bulk, the condensate deviates from perfect homogeneity.

We consider therefore some small deviation from the asymptotic values of the velocity and of the density:
\begin{equation}\label{eq:expansion}
n_{c}=n_{\infty}(1 + 2 u(\xx)), \qquad \vec v = \vec v _{\infty} + \vec w(\xx), \qquad  \textrm{with } u\ll 1, \quad  w\ll v.
\end{equation}
 This implies in particular a rescaling of the speed of sound.
$$ c_{s}^{2} = \frac{\kappa n_{c}}{m} = c_{\infty}^{2} (1 + 2u(\xx)).$$
The acoustic line element \eqref{eq:acousticlineelementasym} becomes then
\begin{equation}
ds^{2}= \frac{n_{c}}{m c_{s}} \frac{m c_{\infty}}{n_{\infty}} \left( -(c_{s}^{2}-v^{2})dt^{2} - 2 v_{i} d\xx^{i}dt + \delta_{ij}d\xx^{i}d\xx^{j} \right),
\end{equation}
where we have introduced a constant prefactor $m c_\infty /n_{\infty}$ in order to have   the conformal factor  asymptotically normalized to one.
Using \eqref{eq:expansion}, together with the coordinate change \eqref{eq:pgtominkowski},  the acoustic line element has the form:
\begin{equation}
ds^{2} = ds^{2}_{\infty} - 3 u(X) c_{\infty}^{2} dT^{2} - 2 w_{i}(X)dT dX^{i} + u(X)\delta_{ij} d X^{i}dX^{j},
\end{equation}
at first order in $u, w_{i}$. Consequently, we see that
\begin{equation}
h_{00}(X) = -3 c_{\infty}^{2}u(X),
\end{equation}
so that  the gravitational field is encoded in the \emph{density perturbation} of the condensate wave-function $\psi$,
\begin{equation}
\Phi_{N}(X) = \frac{3}{2} c_{\infty}^{2} u(X),
\end{equation}
while it is independent from velocity perturbations.

In light of this discussion, we shall discuss situations in which the condensate wavefunction has a constant phase, while its modulus slightly deviates from perfect homogeneity. In order to stay as close as possible to the case of the standard BEC analysis we have just presented, it is convenient to introduce the parametrization:
\begin{equation}\label{perturbation of psi}
\psi = \left(\frac{\mu+\lambda}{\kappa}\right)^{1/2}(1+ u(\xx)),
\end{equation}
where $u(\xx)$ is dimensionless and it is assumed to be very small. In practice, we will assume that it is associated with a localized inhomogeneity of the condensate. At infinity (\ie at the boundary) we  ask that $u\rightarrow0$.
\subsection{Identifying the gravitational potential for the quasi-particles}\label{sec:gravpot}
First of all we want to see if there is a term in the equation of motion for quasi-particles \eqref{dynamics for quasiparticles}, which can be identified as an external potential term. This will allow us to identify the gravitational degrees of freedom.
We have seen in the previous section that they should be encoded in  the condensate wave-function $\psi$. We note however that the dynamics of $\psi$ is essentially non relativistic (\cf \eqref{BdG}). We are therefore looking for a \emph{Newtonian type of gravity}. To identify the Newton potential,  we therefore need to look at the Galilean regime for matter, \ie in the regime where the momentum $p$ of the quasi-particle is such that  $p\ll p_C$ (\cf section \ref{sec:regimes}).

Note that the usual BEC construction given in \eqref{eq:fieldeq0} gives rise to  massless particles. However these latter cannot be handled in the context of Galilean mechanics. It is then not clear at all how one can identify a Poisson like equation for gravity in this case. To solve this issue, we have introduced  $H_\lambda$ in order to generate a non zero mass for the quasi-particle. We can therefore now consider the quasi-particle as a possible source to the Newtonian gravitational potential.

To do identify the Newton potential, we repeat the diagonalization of the Hamiltonian in   \eqref{dynamics for quasiparticles} for the field $\chi$, now including the fluctuations of the condensate.
In this case, the diagonalization procedure is more involved: we have to deal with the non-commuting operators $\nabla^2$ and $u$.
We can not perform it in an exact way. However, we are interested in the Galilean regime for the quasi-particle spectrum, when $p_C\gg p$. It is then a reasonable approximation to neglect all the terms involving the commutators $[\hat{p}^2/2m,u(\xx)]$, which are largely suppressed (with respect to the other terms appearing in the equations) by the mass of the atoms and from the smallness of $u(\xx)$.

With these simplifying assumptions, the Hamiltonian for the quasi-particles in the non-relativistic limit is
\begin{equation}\label{eq:nonrelhquasip}
\hat{H}_{quasip.} \approx \mm \csou^2- \frac{\hbar^2 \nabla^2}{2\mm} + 2\frac{(\mu+\lambda)(\mu+4\lambda)}{\mm \csou^2}u(\xx),
\end{equation}
where the mass of the quasi-particle $\mm$ and for the speed of sound $\csou$ are given in \eqref{eq:soundspeedmass}. We first recognize the constant shift $\mm \csou^2$ of the energy due to the rest mass in the Galilean regime. This term is not affecting the discussion in any ways and can be eliminated. The  term proportional to $u(\xx)$ can be clearly interpreted as an external potential. If we want to  identify it with the gravitational potential $\phigrav$, we need to have
\begin{equation}\label{gravitational potential}
2\frac{(\mu+\lambda)(\mu+4\lambda)}{\mm \csou^2}u(\xx)= \mathcal{M}\phigrav \Leftrightarrow \phigrav(\xx) = \frac{(\mu+4\lambda)(\mu+2\lambda)}{2\lambda m} u(\xx),
\end{equation}
where $\mathcal{M}$ is the mass of the quasi-particles. Note that this identification is  formal, and relies on the way in which the gravitational potential enters the Schroedinger equation for a non-relativistic quantum particle. We should always work with $u$: our definition of $\phigrav$ is dictated from the analogy we want to make with Newtonian gravity. For instance, we see that this definition becomes singular when we deal with massless quasi-particles, \ie when $\lambda\dr 0$.

\subsection{(Modified) Poisson equation}\label{sec:poisson}
Now that we have identified a candidate for the Newton potential $\phigrav$ from the quasi-particles dynamics, we need to check  that it satisfies some sort of Poisson equation. Since the gravitational potential is deduced from  $\psi$ -- as small deviations from perfect homogeneity (\cf \eqref{perturbation of psi}) -- the Poisson equation should be deduced from the BdG equation \eqref{BdG}. With the natural assumption that the potential  is reacting instantaneously to the change of distribution of matter, we can neglect the time derivative and \eqref{BdG} becomes
\begin{equation}\label{BdG1}
\left(\frac{\hbar^{2}}{2m} \nabla^{2} - 2 (\mu+\lambda) \right) u(\xx) = 2 \kappa \left( {\nnn}(\xx)+\frac{1}{2} {\mmm}(\xx) \right)
\end{equation}
We have seen in section \ref{sec:cond} that the terms ${\mmm}(\xx)$ and ${\nnn}(\xx)$ are functions of the atoms $\hat \chi$ outside the condensate  and therefore of the quasi-particle $\hat \phi$, through the Bogoloubov transformation \eqref{bogolobov transf}.  They can be therefore interpreted as the source in the (modified) Poisson equation. We examine now different types of source: either localized particles or plane-waves.

\subsubsection{Localized sources}\label{sec:loca-source}
The most natural source to consider for the Poisson equation  is  a single quasi-particle $\hat{\phi}$ at a given position $\xx_{0}$. However, point-like particles give rise to divergences. We consider therefore   a quasi-particle which is localized around the point $\xx_{0}$, with a non-zero spread to regularize these divergencies. We consider a quasi-particle in a state of the form:
\begin{equation}
\ket{\zeta_{\xx_0}} = \int \dv \xx \zeta_{\xx_0}(\xx) \hat{\phi}^{\dagger}(\xx)\ket{\Omega},\qquad \textrm{with} \qquad \int \dv\xx |\zeta_{\xx_0}(\xx)|^{2} = 1 \Leftrightarrow  \langle{\zeta_{\xx_0}|\zeta_{\xx_0}}\rangle=1.
\end{equation}
 $\zeta_{\xx_0}$ encodes the spreading of the particle around $\xx_{0}$ since
\begin{equation}
\bra{\zeta_{\xx_0}} \hat{\phi}^{\dagger}(\xx) \hphi(\xx) \ket{\zeta_{\xx_0}} = |\zeta_{\xx_0}(\xx)|^{2}.
\end{equation}
We can now determine the value for the  anomalous mass $\mmm$ and anomalous densities $\nnn$ when the quasi-particle is in the state $\ket{\zeta_{\xx_0}}$.
An explicit calculation, given in the appendix \ref{app:source}, gives
\begin{eqnarray} \label{eq:locsource}
\nnn(\xx) = \left| \int \dv \zz f(\xx-\zz) \zeta_{\xx_0}(\zz) \right|^{2}  +\left| \int \dv \zz g(\xx-\zz) \zeta_{\xx_0}(\zz) \right|^{2} + \frac{1}{V} \sum_{\kkk} \beta^{2}(\kkk), \label{eq:anomalousdensity}\\
\mmm(\xx) = 2 \left(\int \dv \zz_{1} g(\xx-\zz_{1})\zeta^{*}_{\xx_0}(\zz_{1})\right) \left( \int \dv \zz_{1} f(\xx-\zz_{2})\zeta_{\xx_0}(\zz_{2})\right)
+ \frac{1}{V} \sum_{\kkk} \alpha(\kkk)\beta(\kkk)\label{eq:anomalousmass},
\end{eqnarray}
where we have introduced the functions $f$, $g$ depending on the Bogoliubov coefficients $\alpha$ and $\beta$
\begin{equation}
f(\xx) = \frac{1}{V} \sum_{\kkk} \alpha(\kkk)e^{i\kkk\cdot \xx}, \qquad g(\xx) = \frac{1}{V} \sum_{k} \beta(\kkk) e^{-i\kkk\cdot \xx}.
\end{equation}
The quantities $\nnn_{\Omega}$ and $\mmm_{\Omega} $  with
\begin{equation}
\nnn_{\Omega} = \frac{1}{V} \sum_{\kkk} \beta^{2}(\kkk), \qquad \mmm_{\Omega} = \frac{1}{V} \sum_{\kkk} \alpha(\kkk)\beta(\kkk),
\end{equation}
are vacuum contributions independent from the presence of actual quasi-particles. They are related to the inequivalence of the particle and quasi-particle vacua, and it can be easily seen that:
\begin{equation}
\nnn_{\Omega} = \bra{\Omega} \hchi^{\dagger}(\xx) \hchi(\xx) \ket{\Omega}, \qquad \mmm_{\Omega} = \bra{\Omega} \hchi(\xx) \hchi(\xx) \ket{\Omega}.
\end{equation}
The functions $f,g$ encode the fact that quasi-particles are collective degrees of freedom and therefore intrinsically some non-local objects. This non-locality is precisely encoded in the Bogoliubov transformation \eqref{bogolobov transf}. Quasi-particles and atoms (\ie local particles) coincide only if we have $\alpha(\kkk)=1, \beta(\kkk)=0$, and therefore $f(\xx) = \delta^{3}(\xx)$,  while $ g(\xx)=0$. Since this is not the case, the anomalous mass and the anomalous density will show an intrinsic non-locality. The spreading characterized by $\ket{\zeta_{\xx_0}}$ encodes some extra non-local effect, \ie the quasi-particle is in some sense an extended object. It was however introduced by hand, for a regularization purpose and therefore is not fundamental as the non-locality introduced by the Bogoliubov transformation.

The  equation \eqref{BdG1} becomes then:
\begin{equation}\label{bdgbis}
\left(\frac{\hbar^{2}}{2m} \nabla^{2} - 2 (\mu+\lambda) \right) u(\xx) = 2 \kappa \left( \tilde{\nnn}(\xx)+\frac{1}{2} \tilde{\mmm}(\xx) \right) + 2 \kappa \left(
\nnn_{\Omega}+\frac{1}{2}{\mmm}_{\Omega} \right),
\end{equation}
 where we have  introduced the quantities
\begin{equation}
\tilde{\nnn}(\xx)=\nnn(\xx)-\nnn_{\Omega}, \qquad \tilde{\mmm}(\xx)=\mmm(\xx)-\mmm_{\Omega},
\end{equation}
which represent the contribution of actual quasi-particles to the anomalous density and anomalous mass, respectively.  By dimensional analysis, the terms $\nnn,\mmm$ have the dimensions of a number density. Since in Newtonian gravity the source for the gravitational field is a mass density, we introduce the mass density distribution:
\begin{equation}
\rho_{\rm matter} (\xx) = \mm\left( \tilde{\nnn}(\xx)+\frac{1}{2} \tilde{\mmm}(\xx) \right).
\end{equation}
With this definition, we can rewrite \eqref{bdgbis} as an equation for the field $\phigrav$:
\begin{equation}\label{eq:modifiedpoisson}
\left( \nabla^{2} -\frac{1}{L^{2}} \right) \phigrav = 4 \pi \gloc \rho_{\rm matter} + \Lambda,
\end{equation}
where we have defined
\begin{eqnarray}
&& \gloc \equiv \frac{\kappa(\mu+4\lambda)(\mu+2\lambda)^{2}}{4 \pi \hbar^{2} m \lambda^{3/2} (\mu+\lambda)^{1/2}} ,\qquad
 \Lambda \equiv \frac{2 \kappa(\mu+4\lambda)(\mu+2\lambda)}{\hbar^{2} \lambda}(\nnn_\Omega +\demi \mmm_\Omega), \label{eq:lambda}\\
&& \label{eq:range}
L^{2} \equiv \frac{\hbar^{2}}{4m(\mu+\lambda)}.
\end{eqnarray}
This particular choice of notation is motivated by the comparison of \eqref{eq:modifiedpoisson} with the Newtonian limit of Einstein equations with a cosmological constant (see, for instance, Eq. (9) of reference \cite{amjp}). For this reason, we can identify these three quantities as the analogous of the Newton constant, the analogous of the cosmological constant and a length scale which represents the range of the interaction, as we are going to discuss below.

To get a better grasp of the physics of the modified Poisson equation \eqref{eq:modifiedpoisson}, we can look at its solution for a given distribution of quasi-particle $\rho_{\rm matter}$.

As it is well known, a solution for the equation
\begin{equation}
\left( \nabla^{2} -\frac{1}{L^{2}} \right) \Phi(\xx) = 4 \pi \gloc \mm \delta^{3}(\xx-\zz),
\end{equation}
is given by the Yukawa potential
\begin{equation}
\Phi_{\rm Y} (\xx; \zz) = \frac{\gloc \mm e^{-|\xx-\zz|/L}}{|\xx-\zz|}.
\end{equation}
On the other hand, a solution for the equation
\begin{equation}
\left( \nabla^{2} -\frac{1}{L^{2}} \right) \Phi(\xx) = \Lambda,
\end{equation}
is just given by the constant solution
\begin{equation}
\Phi_{\Lambda} = - L^{2}\Lambda.
\end{equation}
The linearity of equation \eqref{eq:modifiedpoisson} allows us to use these results to write down a solution for a generic distribution of matter (\ie quasi-particles) as
\begin{equation}
\phigrav(\xx) = \int  \rho_{\rm matter}(\zz) \Phi_{\rm Y}(\xx;\zz) \dv \zz + \Phi_{\Lambda}.
\end{equation}
Solutions of \eqref{eq:modifiedpoisson} are therefore constructed from the Yukawa potential smeared out due to the non-locality of the quasi-particle (with an extra global shift due to the cosmological constant). The Yukawa potential is typically encoding some short range interaction, characterized by the scale $L$ which is  simply related to the healing length  \eqref{healing},
\be \label{eq:rangegrav} L^2=\frac{\xi^2}{2}.\ee
Although this a very short range for the gravitational interaction, this outcome should not come as a surprise.
%but it could have been expected from the very beginning.
In fact, the healing length (\cf \eqref{healing}) characterizes the typical length over which a condensate can adjust to density gradients.  Since density inhomogeneities encode the gravitational interaction, one should expect them to be damped over a distance of the order the healing length.

In the context of relativistic field theory, the short interaction scale for gravity would be translated in a  massive graviton, with mass given by $$M_{grav}^2=\frac{\hbar^2}{L^2\csou^2}=4\frac{\mu+\lambda}{\mu + 2 \lambda} m^{2}.$$
We can then compare the masses of the quasi-particles $\mm$, graviton $M_{grav}$ and atoms $m$,
\begin{equation}
0\leq \mm <m < \sqrt{2}m < M_{grav}\leq 2m,
\end{equation}
which shows the hierarchy of the energy scales present in this system. We notice that the graviton is then always more massive than the quasi-particles, and that this interaction is of very short range, since the $\xi$ is much shorter than the acoustic Compton length\footnote{We are using $c_{sound}$ instead of $c_{light}$ to define all these scales. We have to use the natural units for an hypothetical phononic observer.} of the quasi-particles. In particular, we cannot tune the parameters of the system in such a way to make $M_{grav}$ arbitrarily small, in order to be closer to reality.

\subsubsection{Plane waves}\label{sec:planewave}
While a quasi-particle localized in a given point in space is certainly the most natural source for gravity from the Newtonian perspective, it is interesting also to see what happens when instead we consider quasi-particles with a definite momentum $\pp=\hbar\kkk$.
Let us focus first on the special case of a 1-particle state with momentum $p$, that is $\ket{\pp}= \hat{b}_{\kkk}^{\dagger} \ket{\Omega}$.
The anomalous mass and the anomalous densities become then
\begin{equation}
\nnn(\xx)= \bra{\Omega} \hat{b}^{\dagger} _{\kkk} \hchi^{\dagger}(\xx) \hchi(\xx) \hat{b} _{\kkk} \ket{\Omega},\qquad \mmm(\xx)= \bra{\Omega} \hat{b}^{\dagger} _{\kkk} \hchi(\xx) \hchi(\xx) \hat{b} _{\kkk} \ket{\Omega}.
\end{equation}
To express them in terms of quasi-particles, we need to perform the  Bogoliubov transformations \eqref{bogolobov transf}. As we recalled in the previous section, we can not specify these transformations exactly, due to the presence of the potential $u$.  However, the corrections to the  Bogoliubov coefficients $\alpha(k), \beta(k)$, evaluated in the case $u=0$ in \eqref{Bogoliubovcoeff},   provide corrections to the expressions of $\nnn$ and $\mmm$ which are relevant only beyond the linear order in $u$, which means beyond the approximation we are using. We can then safely neglect these corrections.  Using the Bogoliubov transformation, we obtain explicitely
\begin{eqnarray}
&& \nnn_{k}(x) = \frac{\alpha^{2}(k)+\beta^{2}(k)}{V} + \nnn_\Omega, \qquad
\mmm_{k}(x) = 2 \frac{ \alpha(k) \beta(k)}{V} +  \mmm_{\Omega}. \nn
\end{eqnarray}
where we recognize the contribution of the vacuum  $\nnn_{\Omega}$, $\mmm_{\Omega}$.

The generalization to the case of states containing a definite number of quasi-particles with a given momentum follows in the same way. For these states denoted as $\ket{n(k_1),...,n(k_n)}$, one obtains:
\begin{equation}
\nnn_{k}(x) =\sum_i n(k_i) \frac{\alpha^{2}(k_i)+\beta^{2}(k_i)}{V} + \nnn_\Omega,\qquad
 \mmm_{k}(x) =2  \sum_i n(k_{i})\frac{ \alpha(k_i) \beta(k_i)}{V} +  \mmm_{\Omega}.
\end{equation}
In all these expressions, besides the Bogoliubov coefficients, we recognize the terms $n(k_{i})/V$, which are the number densities of quasi-particles in a given eigenstate of momentum. These number densities, however, are not giving immediately the source term to be inserted into \eqref{eq:modifiedpoisson}, since they are weighted by the Bogoliubov coefficients. This is the representation, in momentum space, of the non-locality we have discussed in position space.

We  rewrite \eqref{eq:modifiedpoisson} with a source term made by a single particle with a given momentum $\pp = \hbar \kkk$.
\begin{equation}\label{eq:modifiedpoissonmomentum}
\left( \nabla^{2} -\frac{1}{L^{2}} \right) \phigrav = 4 \pi G_{N}(k) \rho_{\rm matter} + \Lambda,
\end{equation}
where we have $\rho_{\rm matter} = \mm/V$ since we have just one particle of mass $\mm$, while $\Lambda, L$ are defined as in \eqref{eq:lambda}-\eqref{eq:range}. We encode the effect of the Bogoliubov coefficients, and hence of non-locality, in the ``running'' Newton constant
\begin{equation}
G_{N}(k) = \left(  \alpha^{2}(k)+\beta^{2}(k) +\alpha(k) \beta(k)\right) \gloc.
\end{equation}
The discussion of the solution to this equation is even simpler than the localized state, given that the source term is just constant. Consequently, $\phigrav=const$ is a solution as in the case with purely vacuum contribution. For what concerns the physical effects of this kind of gravitational field, we have to plug this constant solution into \eqref{eq:nonrelhquasip}: this amounts just to a shift of the energy, which is, however, momentum dependent, leading to observable relative energy shifts if different momenta are considered.

%%%%%%%%%%%%%%%%%%%%%%%%%%
\section*{Conclusions and Remarks}
In an analogue gravity model based on a BEC system, the degrees of freedom are separated into  the atoms that condense and the one which do not.  Quasi-particles are then collective degrees of freedom constructed from the un-condensed atoms. The dynamics of the quasi-particles is  encoded, in a given regime, as particles propagating in a curved spacetime metric, which is characterized by the the density $n_c$ and the velocity profile $\vec v$ of the condensate. In this sense, we can expect that gravitational degrees of freedom are encoded in the condensate. Dynamics of the latter is encoded in the BdG equation \eqref{BdG}, which is essentially Galilean. Hence, we can not expect to recover the Einstein equations in this context \cite{Barcelo:2001tb}. Nonetheless, one can still try to interpret  \eqref{BdG} as some sort of Poisson equation for some type of Newtonian gravity.

However, quasi-particles are massless in usual BEC systems and hence they cannot  be considered as sources for the gravitational field in the Poisson equation. We introduced therefore a new term $\hat H_\lambda$ in the dynamics of the BEC which softly breaks the $U(1)$ symmetry and consequently, as we showed in section \ref{sec:quasipart}, generates a mass gap for the quasi-particles. We showed explicitly that the presence of this small symmetry breaking term does not prevent a condensation from happening and still allows a mean field description (which is sufficiently accurate for our purposes). Then, following the usual General Relativity argument, we have argued, in section \eqref{sec:grav},  that the Newtonian potential $\Phi_N$ has to be  related to small inhomogeneities in the condensate density (while perturbations in the velocity profile do not contribute at first order as gravitational degrees of freedom). This conjecture, based on the analysis of a standard BEC system, was then confirmed by a specific analysis of the modified BEC dynamics for an almost homogenous condensate.

The end point of this investigation can be then summarized in the following two equations
\begin{eqnarray}
&& {\vec{F}} = \mm \vec{a} = - \mm \vec{\nabla} \phigrav,\\
&& \left(\nabla^2-\frac{1}{L^2} \right) \phigrav = 4 \pi G_N \rho + \Lambda,\label{poisson}
\end{eqnarray}
where $\mm$ is the mass of the quasi-particle acquired via the soft $U(1)$ symmetry breaking induced by \eqref{sy-br}, $L$ is proportional to the healing length,  $\Lambda$ plays the role of the cosmological constant and $G_N$ is an effective coupling constant that depends on the condensate microphysics and the form of the matter source.

For what regards the latter we have considered two cases: a localized quasi-particle state and a set of plane waves. In the first case the analogue Newton constant is indeed momentum and position independent and the solution of the modified Poisson equation  \eqref{poisson},  has the form of a smeared Yukawa potential. The smearing is due to the fact that quasi-particles are intrinsically non-local objects, being collective degrees of freedom. When considering plane-waves as sources, we have instead that, due to the momentum dependence of the Bogoliubov transformation, $G_N$ is running with the momentum and the solution for the gravitational potential is a constant (albeit a different one for different momenta). One should however be careful: while it is common in quantum field theory (QFT) to encounter the notion of running coupling constants, the origin of the running here is rather peculiar. Indeed, in QFT the running is due to quantum corrections to the tree level/classical action, here the running is due to the inequivalence between the ground state of the Fock spaces of atoms and quasi-particles. Paraphrasing what has been done in the context of emergent geometry, where the notion of ``rainbow geometry" has been introduced, we could speak about ``rainbow dynamics''.

We have also obtained naturally a cosmological constant in the model: vacuum gravitates. It  is induced by the terms $\bra{\Omega}\hchi^{\dagger}\hchi\ket{\Omega},\bra{\Omega}\hchi\hchi\ket{\Omega}$, where $\Omega$ is the state with no quasi-particles. It is entirely due to the (unavoidable) inequivalence between the quasi-particle vacuum and the particle vacuum and cannot be put to zero just tuning the parameters. It represents an interesting alternative to known mechanisms to generate a cosmological constant (see also \cite{Volo} for similar ideas about the nature of the vacuum energy in condensed matter systems).

In conclusion, BEC as an analogue model for gravity presents therefore many differences with a realistic gravity theory. One should not be deceived by this result, as it would have been a preposterous expectation to recover Einstein General Relativity in a Bose Einstein Condensate. The model is however interesting per se as it still encodes a modified Poisson equation and hence provides new insights on the possible origin of the cosmological and Newton constants in emergent gravity scenarios.

As a future development of this investigation, it might be interesting to analyze a 2-BEC model \cite{2-bec}: in fact in this case one could treat a multi-particle system whose richness could allow a closer mimicking of Newtonian gravity. However, the fact that emergent gravity has to be Newtonian in a BEC based analogue model seems to be unavoidable since the gravitational potential depends on the condensate, which is typically described by non-relativistic equations. A possible way to avoid this issue is either to consider relativistic BEC \cite{Rel-BEC} (however in this case we would still expect to get only some type of scalar gravity), or to change completely paradigm and identify gravity not in the condensate but among the perturbations around the condensate (see for example \cite{nordstrom}). We leave these ideas for further investigations.

\acknowledgments{The authors wish to thank C. Barcel\'o, S. Sonego, M. Visser and S. Weinfurtner for useful discussions and comments on the manuscript.}

\appendix

\section{The fluid description}\label{app:fluid}
The Gross--Pitaevskii (GP) equation describing a BEC admits an interesting fluid interpretation, through the Madelung representation.
We are considering the GP equation given in \eqref{eq:GP}
\begin{equation}
i\hbar \frac{\partial}{\partial t} \psi = -\frac{\hbar^{2}}{2m } \nabla^{2} \psi - \mu \psi -\lambda \psi^{*} + \kappa |\psi|^{2} \psi,
\end{equation}
and we want to use the Madelung representation for the complex field $\psi$:
\begin{equation}
\psi = \sqrt{n_{c}} e^{-i\theta/\hbar}.
\end{equation}
When replacing this into the GP equation, dividing by the phase and splitting the resulting expression into the real and imaginary parts we obtain two equations:
\begin{equation}
\dot{n}_{c} + \vec{\nabla} \cdot (n_{c} \vec{v}) = - \frac{\lambda}{\hbar} n_{c} \sin \left( \frac{2\theta}{\hbar} \right),
\end{equation}
\begin{equation}
\dot{\theta} = V_{\mathrm{quantum}} + \frac{m}{2}v^{2} - \mu-\lambda \cos\left( \frac{2\theta}{\hbar} \right) - \kappa n_{c},
\end{equation}
where we have introduced the velocity field $\vec{v}= -\vec{\nabla}\theta/\hbar$, and
\begin{equation}
 V_{\mathrm{quantum}}  = -\frac{1}{\sqrt{n_{c}}} \frac{\hbar^{2}}{2m} \nabla^{2} \sqrt{n_{c}},
\end{equation}
is the familiar quantum potential term.
These two equations, in the case $\lambda=0$, have a nice interpretation as the continuity equation and the Euler equation for a perfect fluid. The continuity equation, in particular, is just the statement about the conservation of the Noether current associated with the $U(1)$ invariance of the system. On the other hand, when $\lambda\neq 0$ the $U(1)$ invariance is broken, and the number operator is no more conserved by the Hamiltonian evolution.

It is interesting to see what happens when we consider the case of homogeneous condensates, $\partial_{\mu} n_{c}=\partial_{\mu} v^{i}=0$. From the first equation we get:
\begin{equation}
\sin \left( \frac{2\theta}{\hbar} \right) = 0 \Leftrightarrow \theta= \frac{l \pi}{2} \hbar, \,\,\,\,\, l\in \mathbb{Z}.
\end{equation}
This result implies that not only $\vec{v}$ is constant, but that actually vanishes. Inserting this result in the second equation we obtain:
\begin{equation}
n_{c} = \frac{\mu + \cos(l\pi) \lambda}{\kappa}.
\end{equation}
From the analysis of the quasi-particle dynamics in section \ref{sec:quasipart}, we have seen that $\lambda<0$ corresponds to a negative mass square, {\ie} tachyonic behavior. Since $\cos(l \pi)=-1$ would be equivalent to changing the sign of $\lambda$, without repeating the analysis of section \ref{sec:becsetup}, we see that $\cos(l\pi)= 1$ is required for stability (no tachyons).
\section{Source term}\label{app:source}
In this section we provide the details of the calculation of the source term for the Poisson equation corresponding to a localized source (\cf \ref{sec:loca-source}).
We have to evaluate the expressions:
\begin{equation}
\nnn(\xx)=\bra{\zeta_{x_0}} \hchi^{\dagger}(\xx) \hchi(\xx) \ket{\zeta_{x_0}}, \qquad \mmm(\xx)=\bra{\zeta_{x_0}} \hchi(\xx) \hchi(\xx) \ket{\zeta_{x_0}},
\end{equation}
where
\begin{equation}
\ket{\zeta_{x_0}} = \int \dv\zz \,\zeta_{x_0}(\xx) \hphi^{\dagger}(\zz) \ket{\Omega}.
\end{equation}
 Let us describe it for $\nnn$, since $\mmm$ can be evaluated following the same steps.
First, one has to write explicitly $\nnn$ in terms of the field operators:
\begin{equation}
\nnn = \int \dv \zz_{1}\dv\zz_{2}\,\zeta_{x_0}^{*}(\zz_{1})\zeta_{x_0}(\zz_{2}) \bra{\Omega} \hphi(\zz_{1})\hchi^{\dagger}(\xx)\hchi(\xx) \hphi^{\dagger}(\zz_{2})\ket{\Omega}.
\end{equation}
Let us evaluate then the expectation value inside the integral. To do this, it is necessary to replace the expansion of the field operators in plane waves, and then to use the Bogoliubov transformation:
\begin{eqnarray}
\bra{\Omega} \hphi(\zz_{1})\hchi^{\dagger}(\xx)\hchi(\xx) \hphi^{\dagger}(\zz_{2})\ket{\Omega} =
\frac{1}{V^{2} }\sum_{\kkk,\kkk',\hhh,\hhh'} e^{i \hhh\cdot \zz_{1}}e^{-i \kkk \cdot \xx}e^{i \kkk'\cdot \xx}e^{-i \hhh'\cdot \zz_{2}}\times  \nonumber\\
\times \bra{\Omega} \hat{b}_{\hhh}(\alpha(\kkk) \hat{b}^{\dagger}_{\kkk} + \beta(\kkk) \hat{b}_{-\kkk})(\alpha(\kkk') \hat{b}_{\kkk'} + \beta(\kkk') \habd_{-\kkk'})
\hat{b}^{\dagger}_{\hhh'} \ket{\Omega}.
\end{eqnarray}
It is easy to see that, in this last expression, there are only two non-vanishing terms
\begin{eqnarray}
\langle v\rangle =\bra{\Omega} \hat{b}_{\hhh}(\alpha(\kkk) \hat{b}^{\dagger}_{\kkk} + \beta(\kkk) \hat{b}_{-\kkk})(\alpha(\kkk') \hat{b}_{\kkk'} + \beta(\kkk') \habd_{-\kkk'})
\hat{b}^{\dagger}_{\hhh'} \ket{\Omega} = \alpha(\kkk)\alpha(\kkk') \bra{\Omega}  \hab_{\hhh} \habd_{\kkk} \hab_{\kkk'} \habd_{\hhh}\ket{\Omega} + \nonumber\\
+\beta(\kkk)\beta(\kkk') \bra{\Omega}  \hab_{\hhh} \hab_{-\kkk} \habd_{-\kkk'} \habd_{\hhh}\ket{\Omega}.
\end{eqnarray}
Using the algebra of the operators $\hab,\habd$, it is easy to see that the expression reduces to:
\begin{equation}
\langle v\rangle = \alpha(\kkk)\alpha(\kkk')\delta_{\hhh,\kkk} \delta_{\hhh',\kkk'} + \beta(\kkk)\beta(\kkk') \left( \delta_{\kkk,\kkk'}\delta_{\hhh,\hhh'} + \delta_{\hhh,-\kkk'}\delta_{\kkk,-\hhh'} \right).
\end{equation}
Consequently,
\begin{equation}
\nnn(\xx) = A(\xx) + B(\xx) + C(\xx),
\end{equation}
where

\begin{equation}
A(\xx) =\frac{1}{V^{2} } \int \dv \zz_{1}\dv\zz_{2}\sum_{\kkk,\kkk',\hhh,\hhh'} e^{i \hhh\cdot \zz_{1}}e^{-i \kkk \cdot \xx}e^{i \kkk'\cdot \xx}e^{-i \hhh'\cdot \zz_{2}}\zeta^{*}_{x_0}(\zz_{1})\zeta_{x_0}(\zz_{2})\alpha(\kkk)\alpha(\kkk')\delta_{\hhh,\kkk} \delta_{\hhh',\kkk'} \;,
\end{equation}
\begin{equation}
B(\xx) =\frac{1}{V^{2} } \int \dv \zz_{1}\dv\zz_{2}\sum_{\kkk,\kkk',\hhh,\hhh'} e^{i \hhh\cdot \zz_{1}}e^{-i \kkk \cdot \xx}e^{i \kkk'\cdot \xx}e^{-i \hhh'\cdot \zz_{2}}\zeta^{*}_{x_0}(\zz_{1})\zeta_{x_0}(\zz_{2})\beta(\kkk)\beta(\kkk')  \delta_{\kkk,\kkk'}\delta_{\hhh,\hhh'}\;,
\end{equation}
\begin{equation}
C(\xx) =\frac{1}{V^{2} } \int \dv \zz_{1}\dv\zz_{2}\sum_{\kkk,\kkk',\hhh,\hhh'} e^{i \hhh\cdot \zz_{1}}e^{-i \kkk \cdot \xx}e^{i \kkk'\cdot \xx}e^{-i \hhh'\cdot \zz_{2}}\zeta^{*}_{x_0}(\zz_{1})\zeta_{x_0}(\zz_{2}) \beta(\kkk)\beta(\kkk')  \delta_{\hhh,-\kkk'}\delta_{\kkk,-\hhh'} \;.
\end{equation}
To manipulate these expression, it is useful to recall the representation of the Dirac delta in a Fourier series:
\begin{equation}
\delta^{3}(\xx_{1}-\xx_{2}) = \frac{1}{V} \sum_{\kkk} e^{-i\kkk(\xx_{1}-\xx_{2})},
\end{equation}
and that the distribution $\zeta_{x_0}$ is normalized,
\begin{equation}
\int \dv \xx |\zeta_{x_0}(\xx)|^{2} =1.
\end{equation}
After straightforward passages we obtain
\begin{equation}
A(\xx)= \left| \int \dv \zz f(\xx-\zz) \zeta_{x_0}(\zz) \right|^{2} ,
\end{equation}
\begin{equation}
B(\xx) = \frac{1}{V} \sum_{\kkk} \beta^{2}(\kkk) ,
\end{equation}
\begin{equation}
C(\xx)= \left| \int \dv \zz g(\xx-\zz) \zeta_{x_0}(\zz) \right|^{2} ,
\end{equation}
and, finally:
\begin{equation}
\nnn(\xx) =\left| \int \dv \zz f(\xx-\zz) \zeta_{x_0}(\zz) \right|^{2}  +\left| \int \dv \zz g(\xx-\zz) \zeta_{x_0}(\zz) \right|^{2} + \frac{1}{V} \sum_{\kkk} \beta^{2}(\kkk)\;.
\end{equation}
where we have introduced the functions:
\begin{equation}
f(\xx) = \frac{1}{V} \sum_{\kkk}  \alpha(\kkk)e^{i\kkk\cdot \xx}, \qquad g(\xx) = \frac{1}{V} \sum_{\kkk} \beta(\kkk) e^{-i\kkk\cdot \xx}.
\end{equation}
Notice that, as a consequence of $\alpha(\kkk)=\alpha(-\kkk),$ $ \beta(\kkk) = \beta(-\kkk)$ and of the fact that these coefficients can be chosen to be real, the functions $f,g$ are real functions.

%%%%%%%%%%%
%%%%%%%%%%%%%
%%%%%%%%%%%%%%%
%%%%%%%%%%%
%%%%%%%%%%%%%
%%%%%%%%%%%%%%%
%%%%%%%%%%%
%%%%%%%%%%%%%
%%%%%%%%%%%%%%%
Applying the same procedure to the term $\mmm(\xx)$ we obtain
\begin{equation}
\mmm(\xx) =2 \left(\int \dv \zz_{1} g(\xx-\zz_{1})\zeta^{*}_{x_0}(\zz_{1})\right) \left( \int \dv \zz_{1} f(\xx-\zz_{2})\zeta_{x_0}(\zz_{2})\right)
+ \frac{1}{V} \sum_{\kkk} \alpha(\kkk)\beta(\kkk)  .
\end{equation}

%%%%%%%%%%%%%%%%%


\begin{thebibliography}{99}


 %\cite{Barcel\'o:2005fc}
\bibitem{Barcelo:2005fc}
  C.~Barcel\'o, S.~Liberati and M.~Visser,
  ``Analogue gravity,''
  Living Rev.\ Rel.\  {\bf 8}, 12 (2005)
  [arXiv:gr-qc/0505065].
  %%CITATION = 00222,8,12;%%

%\cite{Unruh:1980cg}
\bibitem{unruh}
  W.~G.~Unruh,
  ``Experimental black hole evaporation,''
  Phys.\ Rev.\ Lett.\  {\bf 46} (1981) 1351.
  %%CITATION = PRLTA,46,1351;%%

%\cite{Visser:1993ub}
\bibitem{visseracou}
  M.~Visser,
  ``Acoustic propagation in fluids: An Unexpected example of Lorentzian
  geometry,''
  arXiv:gr-qc/9311028.
  %%CITATION = GR-QC/9311028;%%



%\cite{Garay:1999sk}
\bibitem{Garay:1999sk}
  L.~J.~Garay, J.~R.~Anglin, J.~I.~Cirac and P.~Zoller,
  ``Black holes in Bose-Einstein condensates,''
  Phys.\ Rev.\ Lett.\  {\bf 85} (2000) 4643
  [arXiv:gr-qc/0002015];
  %%CITATION = PRLTA,85,4643;%%
  ``Sonic black holes in dilute Bose-Einstein condensates,''
  Phys.\ Rev.\  A {\bf 63} (2001) 023611
  [arXiv:gr-qc/0005131].
  %%CITATION = PHRVA,A63,023611;%%

%\cite{Barcel\'o:2000tg}
\bibitem{Barcelo:2000tg}
  C.~Barcel\'o, S.~Liberati and M.~Visser,
  ``Analog gravity from Bose-Einstein condensates,''
  Class.\ Quant.\ Grav.\  {\bf 18}, 1137 (2001)
  [arXiv:gr-qc/0011026].
  %%CITATION = CQGRD,18,1137;%%

  \bibitem{Balbinot:2004da}
  R.~Balbinot, S.~Fagnocchi, A.~Fabbri and G.~P.~Procopio,
  ``Backreaction in acoustic black holes,''
  Phys.\ Rev.\ Lett.\  {\bf 94}, 161302 (2005)
  [arXiv:gr-qc/0405096].
  %%CITATION = PRLTA,94,161302;%%

  R.~Balbinot, S.~Fagnocchi and A.~Fabbri,
  ``Quantum effects in acoustic black holes: The backreaction,''
  Phys.\ Rev.\  D {\bf 71}, 064019 (2005)
  [arXiv:gr-qc/0405098].
  %%CITATION = PHRVA,D71,064019;%%
%\cite{Balbinot:2004da}

%\cite{Barcel\'o:2007yk}
\bibitem{BLVsemicls collapse}
  C.~Barcel\'o, S.~Liberati, S.~Sonego and M.~Visser,
  ``Fate of gravitational collapse in semiclassical gravity,''
  Phys.\ Rev.\  D {\bf 77} (2008) 044032
  [arXiv:0712.1130 [gr-qc]].
  %%CITATION = PHRVA,D77,044032;%%

%\cite{Barcel\'o:2003et}
\bibitem{BLVFRW}
  C.~Barcel\'o, S.~Liberati and M.~Visser,
  ``Analogue models for FRW cosmologies,''
  Int.\ J.\ Mod.\ Phys.\  D {\bf 12} (2003) 1641
  [arXiv:gr-qc/0305061].
  %%CITATION = IMPAE,D12,1641;%%

%\cite{Weinfurtner:2008if}
\bibitem{Silke-Infla}
  S.~Weinfurtner, P.~Jain, M.~Visser and C.~W.~Gardiner,
  ``Cosmological particle production in emergent rainbow spacetimes,''
  arXiv:0801.2673 [gr-qc].
  %%CITATION = ARXIV:0801.2673;%%

 %\cite{Barcel\'o:2001tb}
\bibitem{Barcelo:2001tb}
  C.~Barcel\'o, M.~Visser and S.~Liberati,
  ``Einstein gravity as an emergent phenomenon?,''
  Int.\ J.\ Mod.\ Phys.\  D {\bf 10} (2001) 799
  [arXiv:gr-qc/0106002].
  %%CITATION = IMPAE,D10,799;%%

\bibitem{blackbook}
C.~J.~Pethick and H.~Smith,
{\em Bose-Einstein Condensation in Dilute Gases,}
Ed. Cambridge University Press, Cambridge, U.K. 2002

\bibitem{castindum}
C.~W.~Gardiner,``A particle-number-conserving Bogoliubov method which demonstrates the validity of the time-dependent Gross--Pitaevskii equation for a highly condensed Bose gas,'' Phys.\ Rev.\ A {\bf 56} (1997), 1414-1423 [arXiv:quant-ph/9703005];


Y.~Castin and R. Dum,
``Low-temperature Bose-Einstein condensates in time-dependent traps: Beyond the $U(1)$ symmetry-breaking approach,''
  Phys.\ Rev. \ A {\bf 57} (1998), 3008-3021.
%\cite{Misner:1974qy}

\bibitem{magnon}
G.~E.~Volovik, ``Phonons in magnon superfluid and symmetry breaking  field",
Pis'ma ZhETF {\bf 87},  736--737 (2008);  JETP Lett. {\bf 87},  
639--640 (2008) [arXiv:0804.3709 [cond-mat.other]].

\bibitem{Misner:1974qy}
  C.~W.~Misner, K.~S.~Thorne and J.~A.~Wheeler,
  ``Gravitation,''
%\href{http://www.slac.stanford.edu/spires/find/hep/www?irn=6627595}{SPIRES entry}
{\it  San Francisco 1973, 1279p}

\bibitem{amjp} Harvey, A., \& Schucking, E.,
``Einstein's mistake and the cosmological constant,''
 Am. J. Phys. {\bf 68} (2000), 723-727.

\bibitem{Volo}
  G.~E.~Volovik,
  "Vacuum energy: Myths and reality,"
  Int.\ J.\ Mod.\ Phys.\  D {\bf 15}, 1987 (2006)
  [arXiv:gr-qc/0604062].
  %%CITATION = IMPAE,D15,1987;%%

\bibitem{2-bec}
 %\cite{Weinfurtner:2006wt}
%\bibitem{Weinfurtner:2006wt}
  S.~Weinfurtner, S.~Liberati and M.~Visser,
  ``Analogue spacetime based on 2-component Bose-Einstein condensates,''
  Lect.\ Notes Phys.\  {\bf 718}, 115 (2007)
  [arXiv:gr-qc/0605121];
  %%CITATION = LNPHA,718,115;%%
%\cite{Liberati:2005pr}
%\bibitem{Liberati:2005pr}
 % S.~Liberati, M.~Visser and S.~Weinfurtner,
  ``Naturalness in emergent spacetime,''
  Phys.\ Rev.\ Lett.\  {\bf 96}, 151301 (2006)
  [arXiv:gr-qc/0512139].
  %%CITATION = PRLTA,96,151301;%%
%\bibitem{Liberati:2005id}
 % S.~Liberati, M.~Visser and S.~Weinfurtner,
  ``Analogue quantum gravity phenomenology from a two-component  Bose-Einstein
  condensate,''
  Class.\ Quant.\ Grav.\  {\bf 23}, 3129 (2006)
  [arXiv:gr-qc/0510125].
  %%CITATION = CQGRD,23,3129;%%
%\cite{Volovik:2006bh}

\bibitem{Rel-BEC}
J. Bernstein and S. Dodelson.
``Relativistic Bose gas"
Phys.\ Rev.\ Lett.\ {\bf 66}, 683 (1991).

\bibitem{nordstrom}
   F.~Girelli, S.~Liberati and L.~Sindoni,
  ``On the emergence of time and gravity,"
  arXiv:0806.4239 [gr-qc].

\end{thebibliography}
\end{document}